\documentclass[aps,prl,twocolumn,showpacs,10pt,superscriptaddress,preprintnumbers]{revtex4-2}
\usepackage{amsmath}
\usepackage{physics}
\usepackage{graphicx}
\usepackage{amssymb}
\usepackage[colorlinks,citecolor=red]{hyperref}

\usepackage{color}    
\usepackage{ulem}
\usepackage{multirow}

\newcommand{\beq}{\begin{equation}}
\newcommand{\eeq}{\end{equation}}
\newcommand{\bea}{\begin{eqnarray}}
\newcommand{\eea}{\end{eqnarray}}
\newcommand{\nn}{\nonumber}

\usepackage{ulem}

\begin{document}

\preprint{
	{\vbox {
			\hbox{\bf MSUHEP-23-001}
}}}
\vspace*{0.2cm}

\title{Discriminating between Higgs Production Mechanisms via Jet Charge at the LHC}

\author{Hai Tao Li}
\email{haitao.li@sdu.edu.cn}
\affiliation{School of Physics, Shandong University, Jinan, Shandong 250100, China}

\author{Bin Yan}
\email{yanbin@ihep.ac.cn (corresponding author)}
\affiliation{Institute of High Energy Physics, Chinese Academy of Sciences, Beijing 100049, China}

\author{C.-P. Yuan}
\email{yuan@pa.msu.edu}
\affiliation{Department of Physics and Astronomy,
Michigan State University, East Lansing, MI 48824, USA}


\begin{abstract}
Discriminating the Higgs production mechanisms plays a crucial role in directly measuring the couplings of Higgs to gauge bosons for probing the nature of the  electroweak symmetry breaking. We propose a novel method to distinguish the Higgs production mechanisms at the LHC by utilizing the jet charge asymmetry of the two leading forward jets in Higgs plus two jets production. This novel observable provides a way to disentangle the $W$-fusion from the $Z$-fusion and gluon fusion processes for the first time, due to the electric charge correlation of the two leading jets in the events. We show that the Higgs couplings to gauge bosons can be well constrained and its conclusion does not depend on the other possible new physics effects which modify the Higgs total or partial width. We also discuss the complementary roles between the proposed jet charge asymmetry measurement and the Higgs signal strength measurements at the HL-LHC in determining the Higgs couplings.
\end{abstract}

\maketitle

\noindent {\bf Introduction:} 
Precisely measuring the interactions between the Higgs boson and the weak gauge bosons ($W$ and $Z$) plays a crucial role to verify the electroweak symmetry breaking (EWSB) mechanism of the Standard Model (SM) and beyond. These couplings have been widely discussed under the framework of $\kappa$ scheme or the Standard Model effective field theory (SMEFT) at the Large Hadron Collider (LHC) and future colliders~\cite{Englert:2014uua,Cheung:2014noa,Bergstrom:2014vla,Falkowski:2015fla,Craig:2015wwr,Corbett:2015ksa,Durieux:2017rsg,DeBlas:2019qco,Cao:2018cms,Chiang:2018fqf,Ellis:2020unq,Araz:2020zyh,Yan:2021tmw,Xie:2021xtl,Bizon:2021rww,Banerjee:2021huv,Sharma:2022epc,Asteriadis:2022ebf}. From the recent global analysis of the ATLAS~\cite{ATLAS:2022vkf} and CMS~\cite{CMS:2022dwd} Collaborations at the 13 TeV LHC, 
the magnitudes of these couplings have been severely constrained by the Higgs data within an uncertainty about $\mathcal{O}(10\%)$, i.e. $\kappa_W=1.05\pm 0.06$, $\kappa_Z=0.99\pm 0.06$ (ATLAS) and $\kappa_W=1.02\pm 0.08$, $\kappa_Z=1.04\pm 0.07$ (CMS), where $\kappa_{W,Z}$ are the effective gauge coupling strengths between the Higgs boson and the $W$ and $Z$ bosons, respectively,
\beq
\mathcal{L}_{hVV}=\kappa_W g_{hWW}^{\rm SM}hW_\mu^+W^{-\mu}+\frac{\kappa_Z}{2}g_{hZZ}^{\rm SM}hZ_\mu Z^\mu,
\eeq
where $g_{hVV}^{\rm SM}=2m_V^2/v$, with $V=W,Z$, are the Higgs couplings to gauge boson $V$ in the SM, and $v=246~{\rm GeV}$ is the vacuum expectation value.

However, to date, all the knowledge of the Higgs couplings is inferred from the global analysis of the Higgs data under the narrow width approximation for the Higgs production and decay. As a result, the measurements of the Higgs properties are strongly depending on the assumption of the Higgs width, which is difficult to be directly measured at the LHC~\cite{ATLAS:2022vkf,CMS:2022dwd}. Therefore, probing the Higgs couplings with the fewest possible theoretical assumptions ({\it e.g.,}  Higgs width) plays a crucial role to verify the nature of the EWSB. One of the approaches to overcome the shortness of the current global analysis is to measure the off-shell Higgs signal strengths~\cite{Caola:2013yja,Campbell:2013wga,Gainer:2014hha,Li:2015jva}. Unfortunately, the Higgs cross sections decrease so fast in off-shell Higgs phase space region that it would be a challenge to measure the Higgs couplings with a high accuracy~\cite{CMS:2019ekd,CMS:2022ley}. Moreover, generally,
the determination of the Higgs couplings also depends on other theoretical assumptions made in the analysis~\cite{Gainer:2014hha,Englert:2014aca,Cacciapaglia:2014rla,Englert:2014ffa,Lee:2018fxj,Cao:2020npb,Azatov:2022kbs}.

\begin{figure}
\centering
\includegraphics[scale=0.32]{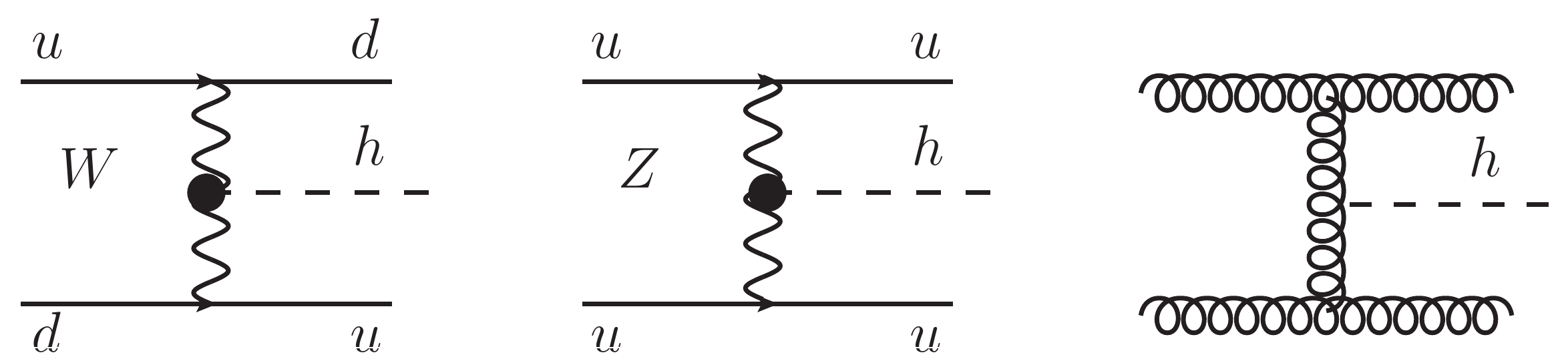}
\caption{Illustrative Feynman diagrams of VBF and GGF Higgs+2 jets production at the LHC. The black dots denote the effective Higgs couplings to gauge bosons. }
\label{Fig:FeyLO}
\end{figure}

For the purpose of determining the $hVV$  couplings, the mechanism of Higgs production via vector boson fusion (VBF) plays an essential role at the LHC. The representative  Feynman diagrams at the leading order are shown in Fig.~\ref{Fig:FeyLO}. 
Since the Higgs+2 jets production through the gluon-gluon fusion (GGF) process is the dominant background for the VBF Higgs production at the LHC, the attempt of separating the VBF from the GGF production has been widely discussed~\cite{Barger:1991ib,Barger:1991ar,Kauer:2000hi,Rentala:2013uaa,Sun:2016mas,Chan:2017zpd,Liu:2019tuy,Sun:2018beb,Chiang:2022lsn}. It shows that the contribution from the GGF process can be largely suppressed by requiring a large rapidity gap and a large invariant mass of the two hardest jets in the Higgs+2 jets events~\cite{Barger:1991ib,Barger:1991ar,Kauer:2000hi}. We can also utilize the different features of soft gluon radiations between the VBF and GGF processes to suppress the contribution from the GGF, {\it e.g.,} the difference in the jet energy profiles~\cite{Rentala:2013uaa}  and the azimuthal angle correlation between the Higgs and  two-jet system~\cite{Sun:2016mas,Sun:2018beb}. But,  none of them could distinguish $WW$ and $ZZ$-boson fusion processes. Separating the $W$ boson's contribution from the VBF Higgs production is an important task for determining $\kappa_{W}$ and $\kappa_Z$, separately, at the LHC.

To suppress the contributions from the $Z$-boson fusion and the GGF processes, we define a novel observable called {\it jet charge asymmetry} between the two leading jets in the Higgs+2 jets production.  For the first time, we demonstrate that the jet charge correlations in Higgs+2 jets production can be used to separate the $W$-boson fusion from the $Z$-boson fusion and GGF processes; the electric charges of the two leading jets in $W$-boson fusion are opposite, while they could be the same or opposite for the $Z$-boson fusion and the GGF processes.
We argue that one could determine the $\kappa_{W,Z}$ without making an assumption on the Higgs width and the other possible new physics effects from Higgs decay. Moreover, the correlation between $\kappa_W$ and $\kappa_Z$, obtained from the jet charges measurement, would be different from other experimental observable ({\it e.g.,} the Higgs signal strength measurement) in the VBF processes. Therefore, the jet charge correlation in the production of  Higgs+2 jets could play an important, and complementary, role in determining the $hVV$ couplings.

\vspace{3mm}
\noindent {\bf  Jet charge:~}%
Jet charge can be used to mimic the electric charge of the parent parton which evolves into a collimated spray of particles. It is defined as a  weighted sum of the electric charge of the jet constituents~\cite{Field:1977fa,Krohn:2012fg,Waalewijn:2012sv},
\beq
Q_J=\frac{1}{(p_T^J)^\kappa}\sum_{i\in {\rm jet}}Q_i(p_T^i)^\kappa,\quad \kappa>0,
\eeq
with $p_T^J$  the transverse momentum of the jet,  $p_T^i$ and $Q_i$ are the transverse momentum and electric charge of particle $i$ inside the jet.  And $\kappa$ is a free parameter which suppresses  the contribution from soft radiations.  
The theoretical framework to calculate  jet charge in QCD was proposed in Refs.~\cite{Krohn:2012fg,Waalewijn:2012sv}. As one of the jet substructure techniques, jet charge has been used for quark/gluon jet discrimination~\cite{Fraser:2018ieu,Larkoski:2019nwj,Gianelle:2022unu}. There are many efforts using jet charge as a tool to tag the hard process~\cite{Chen:2019uar} and to search for new physics signals~\cite{Li:2021uww,Wong:2023vpx}.  Jet charge has also been applied to probe nuclear  medium effects in heavy-ion and electron-ion collisions~\cite{Chen:2019gqo,Li:2019dre,Li:2020rqj}, as well as quark flavor structure of the  nucleon~\cite{Kang:2020fka,Kang:2021ryr,Lee:2022kdn}. 
Experimentally, jet charge has been recently measured by ATLAS and CMS Collaborations~\cite{CMS:2014rsx,ATLAS:2015rlw,CMS:2017yer,CMS:2020plq}.
The theoretical and experimental efforts show that the jet charge serves well for identifying the charge of the primordial parton of the hard scattering. In this letter, we will apply  the jet charge observable in the production of Higgs+2 jets to discriminate the Higgs production mechanisms at the LHC and to further constrain the Higgs couplings to gauge bosons.

\vspace{3mm}
 \noindent{\bf Collider simulation:} 
We perform a detailed Monte Carlo simulation to explore the potential of the LHC for discriminating Higgs production mechanisms and probing the $hVV$ couplings via jet charge correlations.
We generate both the $W/Z$ fusion and GGF Higgs production processes at the 14 TeV LHC by MadGraph5~\cite{Alwall:2014hca} at the parton-level with the CT14LO parton distribution functions (PDFs)~\cite{Dulat:2015mca}. The following basic cuts for the jets are required in the partonic final state: $p_T^j>20~{\rm GeV}$ with $|\eta_j|<5$, where $p_T^j$ and $\eta_j$ denotes the transverse momentum and rapidity of jet, respectively. The GGF process was generated under the heavy top quark limit, {\it i.e.,} the EFT operator $\alpha_s/(12\pi v)hG_{\mu\nu}^aG^{\mu\nu,a}$, which provides a good approximation when the top quarks inside the loop are off-shell. To get the isolated jets, we require the cone distance between the two jets $\Delta R>0.4$. To suppress the contribution from the GGF process, we further require the invariant mass of jet pair $m_{jj}>110~{\rm GeV}$ and rapidity gap between the two jets $|\Delta\eta_{jj}|>2.5$. The jet is defined based on the  anti-$k_T$ algorithm~\cite{Cacciari:2008gp} with the radius parameter $R=0.4$.  Then we pass the events to PYTHIA~\cite{Sjostrand:2014zea,Bierlich:2022pfr} for parton showering and hadronization.  At the analysis level, all the events are required to pass a set of selection cuts following the settings from Ref.~\cite{ATLAS:2022vkf},
\begin{align}
&p_T^j>30~{\rm GeV}, & &1<|\eta_j|<4.5, \nn\\
& m_{jj}>120~{\rm GeV},& &|\Delta\eta_{jj}|>3.5,&& |\eta_h|<2.5\, ,
\label{eq:cuts}
\end{align}
where $\eta_h$ is the pseudorapidity of the Higgs boson. 

\begin{figure*}
\centering
\includegraphics[scale=0.27]{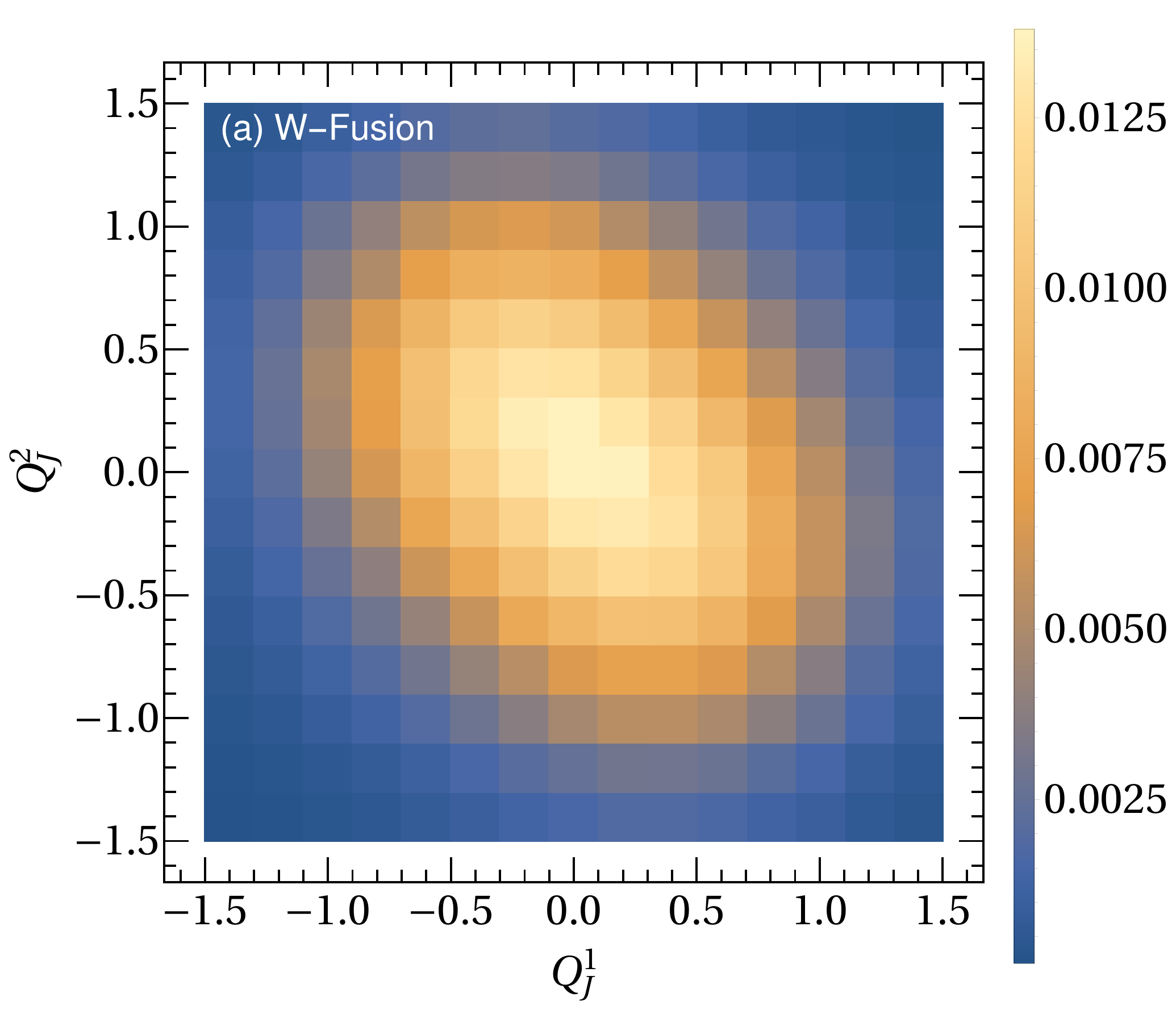}
\includegraphics[scale=0.27]{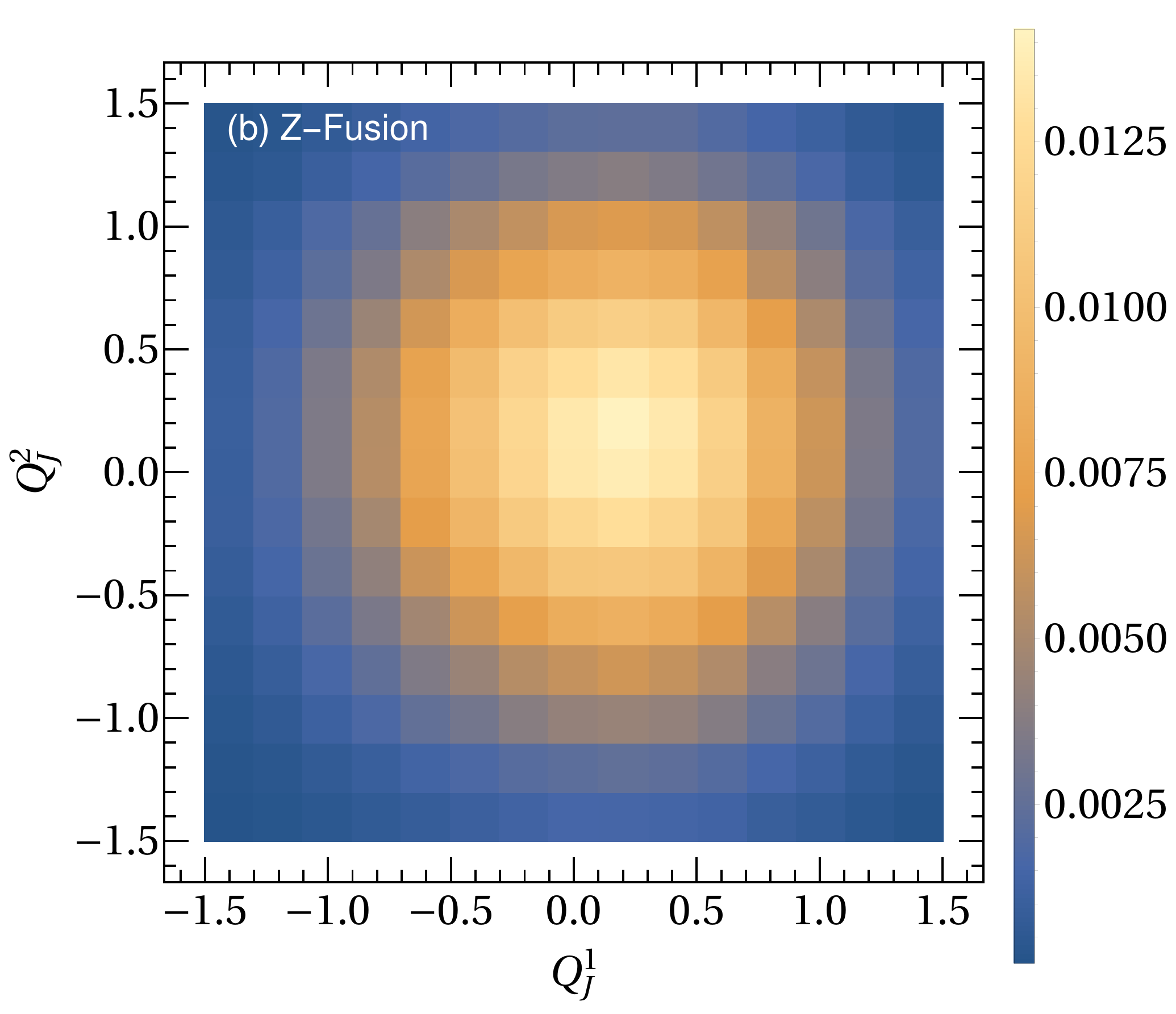}
\includegraphics[scale=0.27]{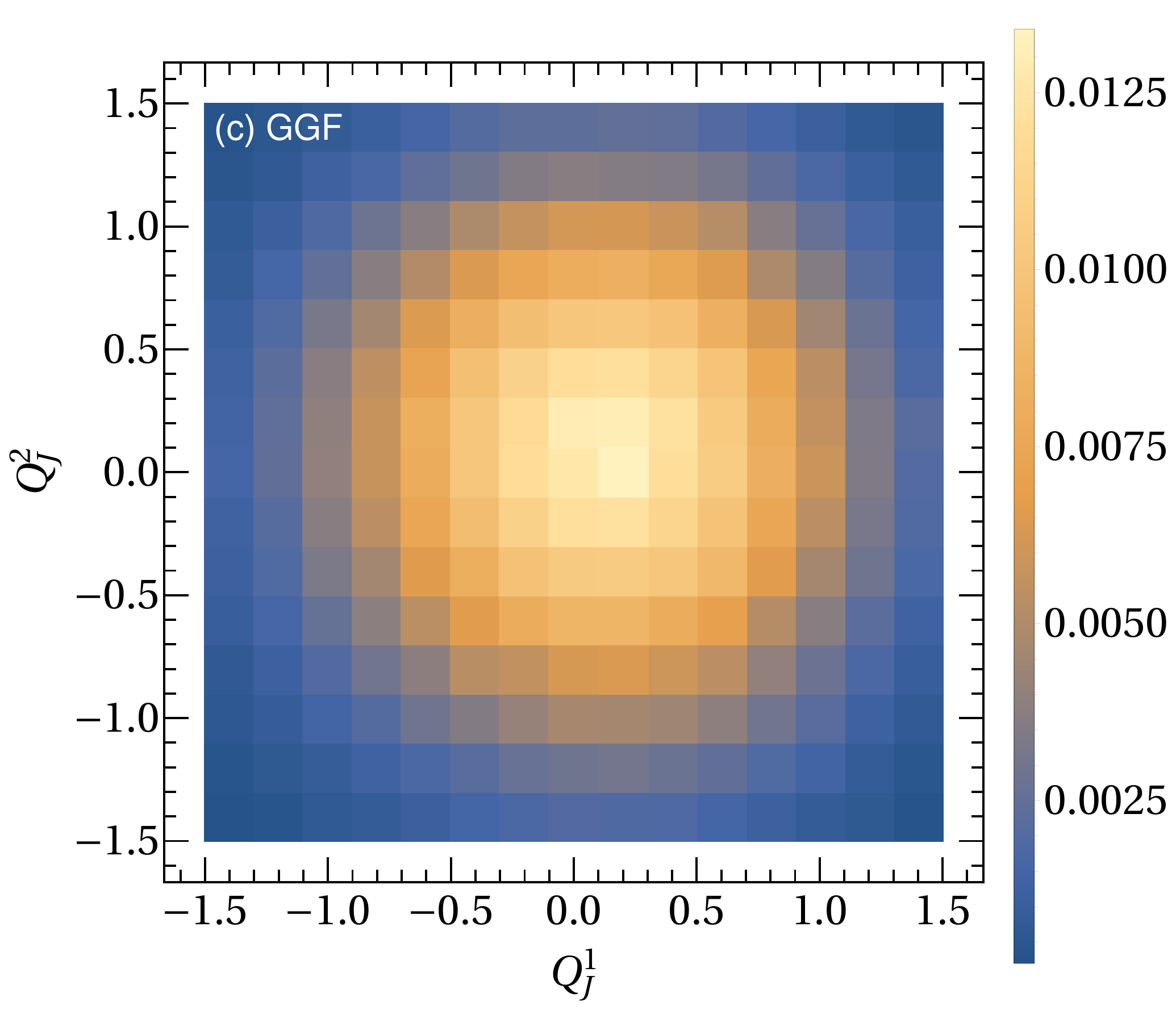}
\caption{The jet charge correlations between the leading and subleading jets from $W$-fusion, $Z$-fusion, and GGF processes with jet charge parameter $\kappa=0.3$.}
\label{Fig:jetcharge}
\end{figure*}

\begin{figure}
\centering
\includegraphics[scale=0.5]{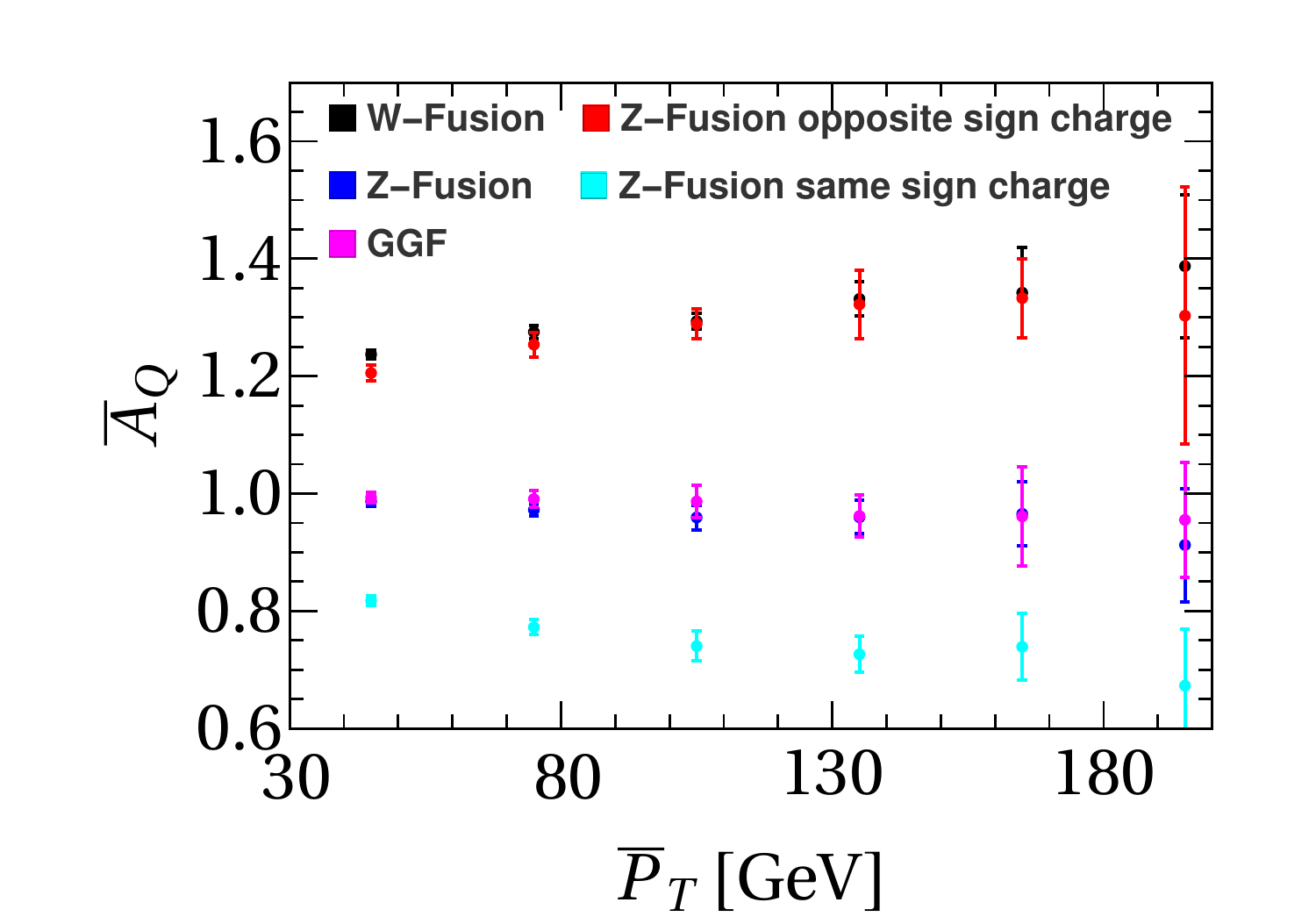}
\caption{The jet charge asymmetry $\overline{A}_Q$ distributions as a function of the average transverse momentum of the two leading jets $\overline{p}_T$ at the 14 TeV LHC with the integrated luminosity $\mathcal{L}=300~{\rm fb}^{-1}$. Its systematic uncertainties are assumed to cancel out.}
\label{Fig:asy}
\end{figure}

On average the sign of the jet charge is consistent with the charge of the parton which evolves into the jet from the measurements at the LHC~\cite{ATLAS:2015rlw,CMS:2017yer}. Therefore, the charge correlations for different VBF channels indicate the charge correlations of the parton in the hard process. Figure~\ref{Fig:jetcharge} shows the jet charge correlations between the leading and subleading jets from the $W$-fusion, $Z$-fusion and GGF processes after the kinematic cuts (see Eq.~\eqref{eq:cuts}) with the benchmark jet charge parameter $\kappa=0.3$. It clearly shows that the opposite sign of the two leading jet charges is favored in $W$-fusion process due to the partonic nature of the hard scattering at the leading order. However, this feature is disappeared in $Z$-fusion and GGF processes. It arises from the fact that both the opposite and same sign electric charges of the two leading jets in $Z$-fusion could be generated with a comparable production rate, while the  sign of the jet charges in the GGF process is arbitrary. Motivated by this, we define the jet charge asymmetry of the two leading jets in Higgs+2 jets production by
\beq
A_Q=\left|\frac{Q_J^1-Q_J^2}{Q_J^1+Q_J^2}\right|,
\label{eq:asy}
\eeq
where $Q_J^{1,2}$ is the jet charge of the leading and subleading jets, respectively.
The advantage of this observable is that the systematic uncertainties, which are the dominant errors in jet charge measurements at the ATLAS and CMS Collaborations~\cite{ATLAS:2015rlw,CMS:2017yer}, are expected to be canceled.
This conclusion has been verified by the previous studies of other track-based observable in the HERA measurement~\cite{H1:2009lef}.
On the other hand, the jet charge distributions are sensitive only to the weight of the jet flavor of final states in the hard scattering but not to the Higgs decay information ({\it i.e.,} the Higgs width and the other possible new physics effects in Higgs decay). However, the jet charge asymmetry in Eq.~\eqref{eq:asy} may be unstable since it can be much enhanced for the events with $Q_J^1+Q_J^2\sim 0$,
therefore, we will utilize the average values of the jet charges to define this asymmetry in this paper, i.e.,
\beq
\overline{A}_Q\equiv\frac{\langle \left|Q_J^1-Q_J^2\right |\rangle}{\langle \left|Q_J^1+Q_J^2\right |\rangle} \equiv \frac{\langle Q^{(-)} \rangle  }{\langle Q^{(+)} \rangle},
\eeq 
where $\langle Q\rangle$ denotes the average value of the quantity $Q$ and $Q^{(\pm)}=|Q_J^1\pm Q_J^2|$.
From the definition, if the charges have the opposite sign such as the partonic process for $W$-fusion case, the asymmetry is expected to be  $\overline{A}_Q>1$. However, if the charges have the same sign,  {\it e.g.,} the $Z$-fusion process $uu\to u u h$,  $\overline{A}_Q<1$. For the GGF process the charges are fully uncorrelated so that  $\overline{A}_Q\sim 1$. 

Figure~\ref{Fig:asy} shows the jet charge asymmetry $\overline{A}_Q$ as a function of the average transverse momentum of the two leading jets ($\overline{p}_T\equiv (p_T^1+p_T^2)/2$) in Higgs+2 jets production at the 14 TeV LHC with the integrated luminosity $\mathcal{L}=300~{\rm fb}^{-1}$. The statistical uncertainties are estimated from the Pythia simulation by generating a large number of pseudo experiments. We assume that the statistical errors follow the Gaussian distribution and could be rescaled to any integrated luminosity by the event numbers. We have checked that $\overline{A}_Q$ does not noticeably depend on the choice of $\kappa$ value, when varying from 0.3 to 0.7.
As expected we find  $\overline{A}_Q>1$ for the $W$-fusion (black points) since the sign of the jet charges are opposite. However, the asymmetry is close to one for the $Z$-fusion as shown with blue points in Fig.~\ref{Fig:asy}. To further understand the behavior of the $Z$-fusion, we also show $\overline{A}_Q$ for the $Z$-fusion with opposite (red points) and same sign jet charges (cyan points) in the same figure and the results are consistent with the argument before. 
When we combine all possible channels, this asymmetry would be close to one for the $Z$-fusion process since the opposite and same sign jet charge configurations could give a comparable contribution.
The dominant contributions for the GGF process come from the $qg$ and $gg$ initial states, in result, we could expect that $\overline{A}_Q\sim 1$ due to the uncorrelated nature of the jet charges in this process.

\begin{figure}
\centering
\includegraphics[scale=0.33]{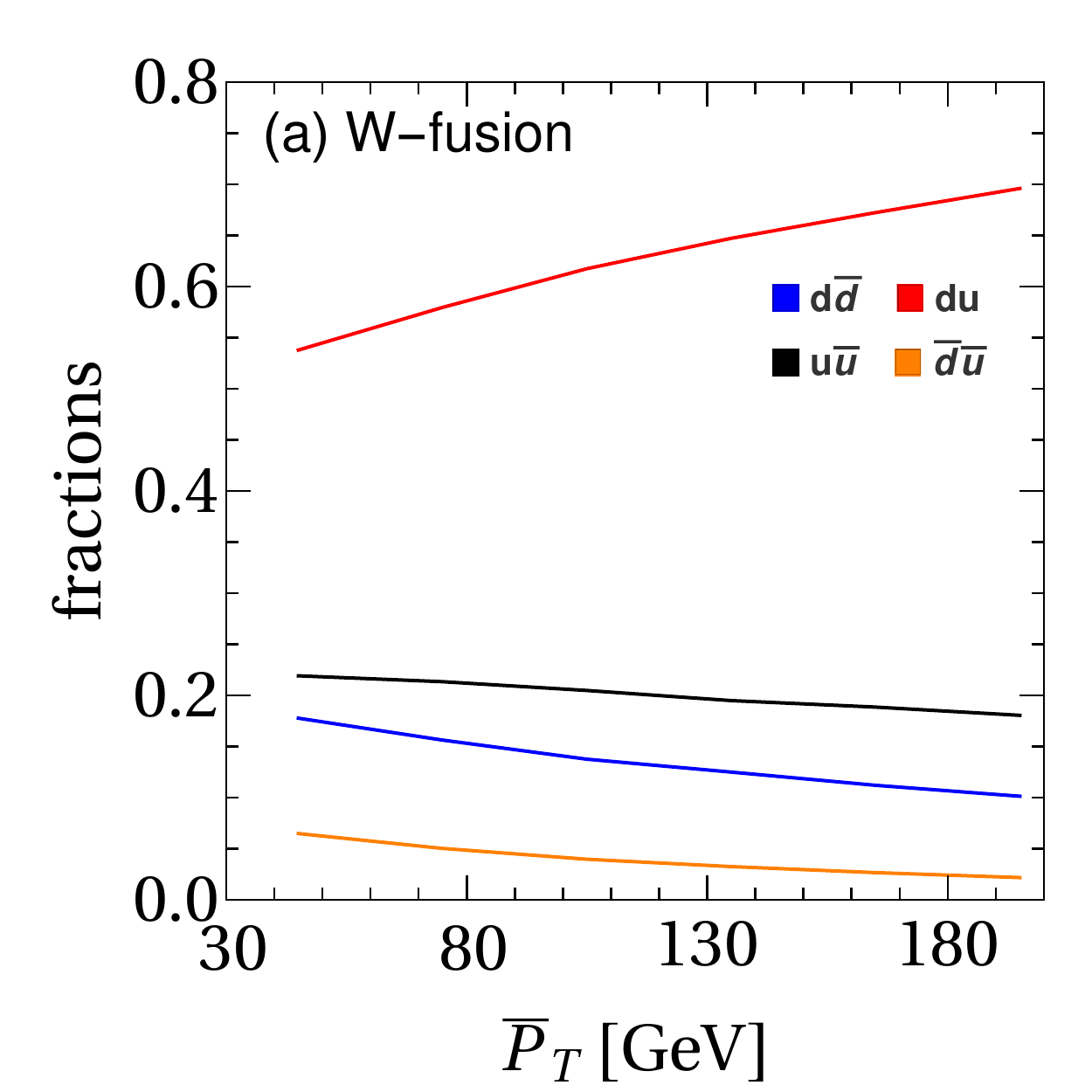}
\includegraphics[scale=0.33]{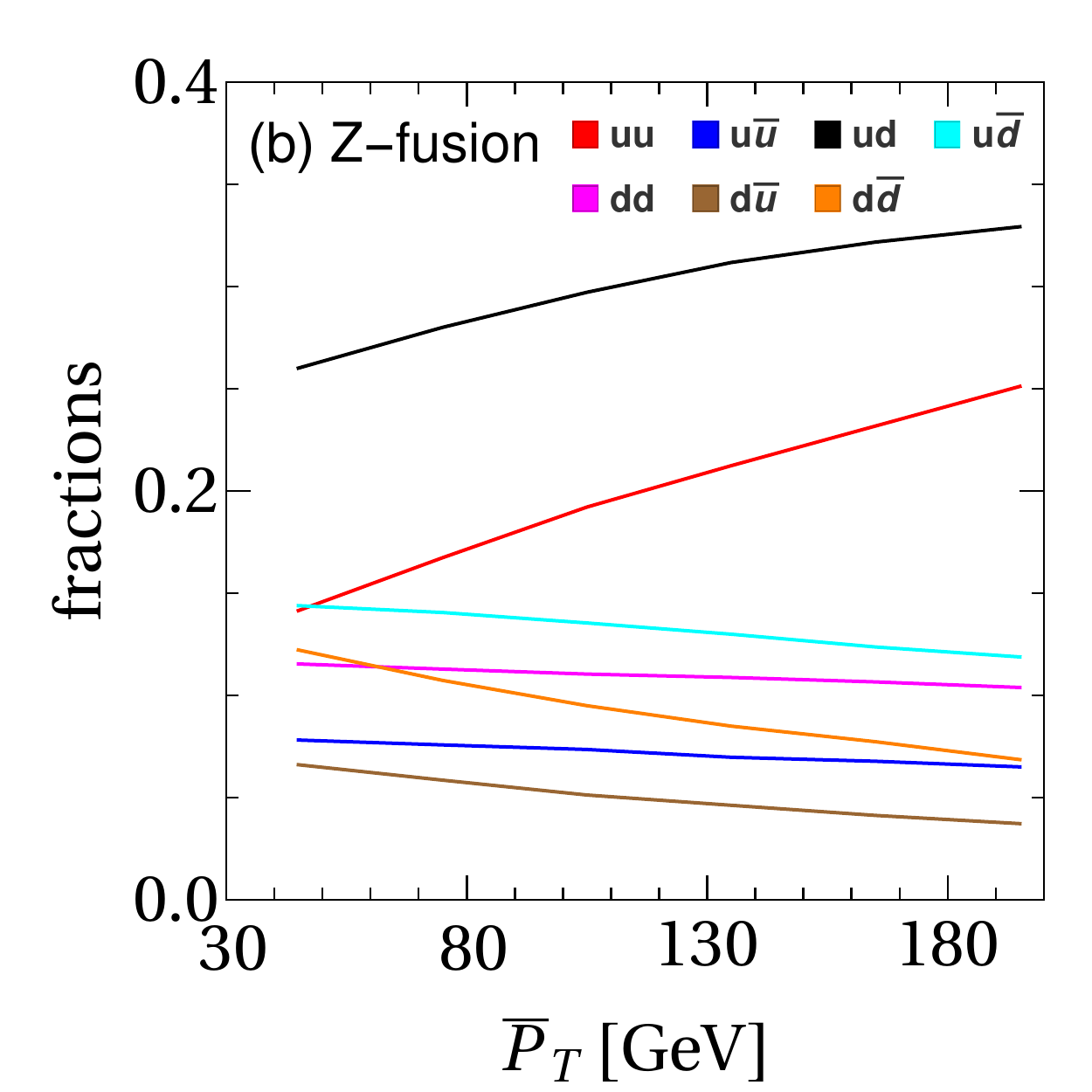}
\caption{Various initial state ($qq^\prime$) fraction distributions $f_{qq^\prime}$ for Higgs+2 jets production via  $W$-fusion and $Z$-fusion at the 14 TeV LHC, as a function of the average jet $\overline{p}_T$.
}
\label{Fig:frac}
\end{figure}

We also notice that the jet charge asymmetry is not very sensitive to $\overline{p}_T$ for the $Z$-fusion and the GGF processes. In particular, it exhibits a weak $\overline{p}_T$ dependence for the $W$-fusion and the $Z$-fusion with opposite or same sign jet charges. To be clear about the $\overline{p}_T$ dependence for the $W$ and the $Z$-fusion processes, we calculate various initial state ($qq^\prime$) fraction distributions $f_{qq^\prime}$ for the $W$-fusion and the $Z$-fusion after the kinematic cuts in Eq.~\eqref{eq:cuts} at the leading order; see Fig.~\ref{Fig:frac}.  
It  shows that the $du$ quark initial state dominates the cross section in $W$-fusion (see Fig.~\ref{Fig:frac}(a)) and it will induce a weak $\overline{p}_T$ dependence for the charge asymmetry in Fig.~\ref{Fig:asy} (black points). For the $Z$-fusion process, we only show the fractions of the dominated channels in Fig.~\ref{Fig:frac}(b). The weak $\overline{p}_T$ dependence for the charge asymmetry of the $Z$-fusion with opposite (red points in Fig.~\ref{Fig:asy}) and same sign (cyan points in Fig.~\ref{Fig:asy}) jet charges can also be understood from the behavior of the dominated fractions $f_{ud}$ and $f_{uu}$.  As a consequence, the jet charge asymmetry of the $Z$-fusion will not be sensitive to the $\overline{p}_T$ after we combine all the subprocesses.

Next, we utilize the $\overline{A}_Q$ information to constrain the $hVV$ couplings. After combing the $W$-fusion, $Z$-fusion and GGF cross sections,  the jet charge asymmetry in Higgs+2 jets process reads as 
\begin{align}
\overline{A}_Q^{\rm tot}=
    & \frac{f_W \langle Q^{(-)}\rangle_W +f_Z \langle Q^{(-)}\rangle_Z +f_G \langle Q^{(-)}\rangle_G }{f_W \langle Q^{(+)}\rangle_W+f_Z \langle Q^{(+)}\rangle_Z +f_G \langle Q^{(+)}\rangle_G } \;,
\end{align}
where the subscripts $W$, $Z$ and $G$ represent the contribution from  different channels accordingly. The fractions $f_{i}(\overline{p}_T,\kappa_{W/Z})$ is defined as $\sigma_i/(\sigma_W+\sigma_Z+\sigma_G )$ with $i=W,Z,G$.  
It shows that the $W$-boson fusion dominates the cross section after applying the kinematic cuts in Eq.~\eqref{eq:cuts}. The contamination from GGF can be further suppressed with boosted decision trees analysis, as shown in Fig. 2 of Ref.~\cite{ATLAS:2022tan}.
To estimate the impact of the jet charge asymmetry on constraining the $hVV$ couplings, we consider $h\to WW^*(\to 2\ell2\nu_{\ell}) {\, \rm and \,} ZZ^*(\to 4\ell)$, with $\ell=e,\mu,\tau$, as our benchmark decay processes in this work. The dominant background with these decay products is the GGF process~\cite{ATL-PHYS-PUB-2016-008}. But we emphasize that the correlations between $\kappa_W$ and $\kappa_Z$ measured from jet charges would not be sensitive to the Higgs decay information, which has been canceled in the definition of fractions. It can only change the statistical uncertainties in each $\overline{p}_T$ bin through the event numbers. The result could potentially be improved if we combine more decay modes of the Higgs boson, which is however beyond the scope of this work. Performing the pseudo experiments,  we conduct a combined $\chi^2$ analysis as
\beq
\chi^2=\sum_i\left[\frac{\overline{A}_Q^{i,\rm tot}-\overline{A}_Q^{i,\rm tot,0}}{\delta \overline{A}_Q^{i,\rm tot}}\right]^2,
\label{eq:chi2}
\eeq
where $\overline{A}_Q^{i,\rm tot,0}$ and $\delta\overline{A}_Q^{i,\rm tot}$ are the jet charge asymmetry in the SM ({\it i.e.,} $\kappa_{W}=\kappa_{Z}=1$) and the statistical uncertainty of the $i$-th bin, respectively. For simplicity, we have assumed that the experimental values of jet charges agree with the SM predictions. Note that we have rescaled the statistical uncertainties to include the branching ratio ${\rm BR}(h\to 2\ell2\nu_{\ell}/4\ell)$ in each bin in our $\chi^2$ analysis. The kinematic cuts for the decay products of Higgs boson can also change the total event numbers. However, such effects should not significantly change the conclusions of this paper and will be ignored in the following analysis. It shows that the typical error of $\overline{A}_Q^{\rm tot}$ is around $\mathcal{O}(1\%)$, which strongly depends on the average jet $\overline{p}_T$.

\begin{figure}
\centering
\includegraphics[scale=0.45]{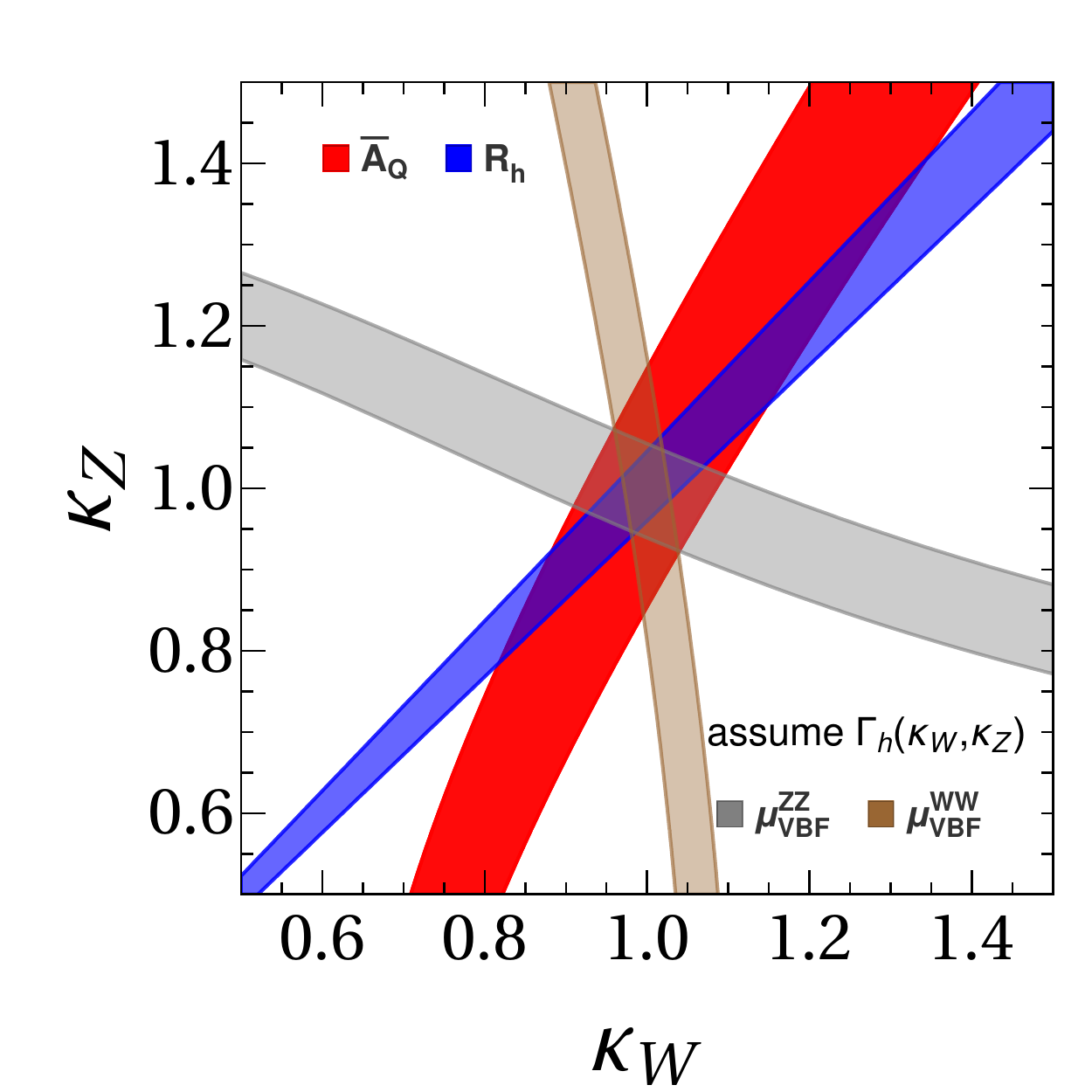}
\caption{The expected constraints at the HL-LHC on $\kappa_{W,Z}$ from the jet charge asymmetry $\overline{A}_Q$ (red). The blue region is bounded by the $R_h$ in $gg\to h$ production. The gray and brown regions represent the constraints from the VBF Higgs signal strength measurements via the decay modes $h\to ZZ^*$ and $h\to WW^*$, respectively, and they depend on the assumption of the Higgs boson width $\Gamma_h$. }
\label{Fig:Limit}
\end{figure}

In Fig.~\ref{Fig:Limit}, we show the expected limits on the $hVV$ couplings at the 68\% confidence level (C.L.) for the integrated luminosity $3000~{\rm fb}^{-1}$, obtained from the jet charge asymmetries (red band) at the 14 TeV LHC (HL-LHC). 
It is evident that $\kappa_{W}$ and $\kappa_{Z}$ cannot be uniquely determined by the measurement of jet charge asymmetry alone. 
In order to further reduce the error band in the parameter space of $\kappa_{W}$ and $\kappa_Z$, without making any assumption on the total decay width ($\Gamma_h$) of the Higgs boson, we consider the following ratio of the Higgs signal strength measurements,   
\beq
R_h\equiv\frac{\mu(gg\to h\to WW^*)}{\mu(gg\to h\to ZZ^*)}=\frac{\kappa_W^2}{\kappa_Z^2} \, ,
\eeq
where the signal strength modifier $\mu(gg\to h \to VV^*)$ is defined as the measured cross section relative to the SM expectation. Hence, any new physics contributions to $gg \to h$ production cross section would cancel in $R_h$.
Its constraint on $\kappa_{W}$ and $\kappa_Z$ at the HL-LHC is shown as the blue band in Fig.~\ref{Fig:Limit}, where the error of $R_h$ is taken to be 5.7\%~\cite{Cepeda:2019klc}.
It is evident that the combined analysis of the  $\overline{A}_Q^{\rm tot}$ and $R_h$ data can further constrain the allowed $\kappa_{W}$ and $\kappa_Z$ values at the HL-LHC.  
We note that the correlation between $\kappa_W$ and $\kappa_Z$, extracted from  $\overline{A}_Q^{\rm tot}$, is dependent of the fraction ($f_{W,Z}$) of production cross section contributed by various subprocesses, which is sensitive to the kinematic selection of the VBF events.
Hence, it is possible to apply some advanced technologies, such as Boosted Decision Tree or Machine Learning, to further improve the constraints on  $\kappa_{W}$ and $\kappa_Z$, which is however beyond the scope of this work.

For completeness, we also show in Fig.~\ref{Fig:Limit} the expected limits from the Higgs signal strength measurements of the VBF Higgs production with $h\to WW^*$ (denoted by $\mu_{\rm VBF}^{WW}$ in the figure, brown region) and $h\to ZZ^*$ (denoted by $\mu_{\rm VBF}^{ZZ}$ in the figure, gray region) at the HL-LHC~\cite{Cepeda:2019klc}.
 Here, the error of the signal strength of the VBF production with $h\to WW^*$ and $h\to ZZ^*$ is taken to be 5.5\% and 9.5\%, respectively~\cite{Cepeda:2019klc}.
We note that while the constraints imposed by the 
$\overline{A}_Q^{\rm tot}$ and $R_h$ data are independent of $\Gamma_h$, those imposed by  $\mu_{\rm VBF}^{WW}$ and $\mu_{\rm VBF}^{ZZ}$ are dependent of the assumption made in the calculation of $\Gamma_h$. Here, we assumed that $\Gamma_h$ is only modified by the values of $\kappa_{W}$ and $\kappa_{Z}$.

\vspace{3mm}
\noindent {\bf Conclusions:~}%
In this Letter, we propose a novel and feasible method to separate the $W$-fusion from the $Z$-fusion and the gluon-fusion (GGF) processes by utilizing the jet charge asymmetry $\overline{A}_Q$ of the two leading jets in Higgs+2 jets production at the HL-LHC. This is crucial for separately measuring the couplings of Higgs boson to $W^+W^-$ and $ZZ$ gauge bosons, without the need of making any assumption about the decay width of Higgs boson. Owing to the partonic nature of the hard scattering, we demonstrate that the asymmetry $\overline{A}_Q>1$ for $W$-fusion, while it is always close to one for the $Z$-fusion and the GGF processes. Such a different feature can be used to discriminate the Higgs production mechanisms and to determine the Higgs couplings to the gauge bosons $\kappa_{V}$.
While the usual methods of determining $\kappa_{V}$ from the Higgs signal strength measurement depends on the assumption of $\Gamma_h$, the proposed measurement of $\overline{A}_Q^{\rm tot}$ does not rely on such an assumption.

\vspace{3mm}
\noindent{\bf Acknowledgments.}
HTL is supported  by the National Science Foundation of China under grant  No. 12275156. 
BY is supported by the IHEP under Contract No. E25153U1. CPY is supported by the U.S.~National Science Foundation under Grant No.~PHY-2013791 and is grateful for the support from the Wu-Ki Tung endowed chair in particle physics.

\bibliographystyle{apsrev}
\bibliography{reference}

\end{document}